\documentclass[9pt, aps, prl, twocolumn, showpacs, superscriptaddress, notitlepage]{revtex4-1}

\usepackage[utf8]{inputenc} 
\usepackage[T1]{fontenc} 
\usepackage[english]{babel} 
\usepackage{lmodern} 
\usepackage{geometry} 
\usepackage{setspace} 
\usepackage{indentfirst} 
\usepackage{graphicx} 
\usepackage{hyperref} 
\usepackage{natbib} 
\usepackage{verbatim} 
\usepackage[pangram]{blindtext} 

\usepackage{amsmath, amsthm, amscd, amssymb} 
\usepackage[all,cmtip]{xy} 
\usepackage{listings} 
\usepackage{physics} 

\def \dif {\mathrm{d}}

\newcommand{\odr}{\omega_{\rm dr}}

\makeindex

\begin{document}
\title{Optimal Feedback Cooling of a Charged Levitated Nanoparticle with Adaptive Control}

\author{Gerard P. Conangla}\email{gerard.planes@icfo.eu}
\affiliation{ICFO Institut de Ciencies Fotoniques, Mediterranean Technology Park, 08860 Castelldefels (Barcelona), Spain}

\author{Francesco Ricci}
\affiliation{ICFO Institut de Ciencies Fotoniques, Mediterranean Technology Park, 08860 Castelldefels (Barcelona), Spain}

\author{Marc T. Cuairan}
\affiliation{ICFO Institut de Ciencies Fotoniques, Mediterranean Technology Park, 08860 Castelldefels (Barcelona), Spain}

\author{Andreas W. Schell}
\affiliation{ICFO Institut de Ciencies Fotoniques, Mediterranean Technology Park, 08860 Castelldefels (Barcelona), Spain}
\affiliation{Quantum Optical Technology Group, Central  European  Institute  of Technology,  Brno  University  of  Technology, 612  00  Brno,  Czech  Republic}

\author{Nadine Meyer}\email{nadine.meyer@icfo.eu}
\affiliation{ICFO Institut de Ciencies Fotoniques, Mediterranean Technology Park, 08860 Castelldefels (Barcelona), Spain}

\author{Romain Quidant}\email{romain.quidant@icfo.eu}
\affiliation{ICFO Institut de Ciencies Fotoniques, Mediterranean Technology Park, 08860 Castelldefels (Barcelona), Spain}
\affiliation{ICREA-Institució Catalana de Recerca i Estudis Avançats, 08010 Barcelona, Spain}

\date{\today}

\begin{abstract}
We use an optimal control protocol to cool one mode of the center of mass motion of an optically levitated nanoparticle. The feedback technique relies on exerting a Coulomb force on a charged particle with a pair of electrodes and follows the control law of a linear quadratic regulator, whose gains are optimized by a machine learning algorithm in under 5 s. With a simpler and more robust setup than optical feedback schemes, we achieve a minimum center of mass temperature of 5 mK at $3\times 10^{-7}$ mbar and transients 10 to 600 times faster than cold damping. This cooling technique can be easily extended to 3D cooling and is particularly relevant for studies demanding high repetition rates and force sensing experiments with levitated objects.
\end{abstract}

\maketitle

\textbf{Introduction}. With the recent Nobel prizes for the detection of gravitational waves\citep{ligo2017} and optical tweezers\citep{ashkin1980applications, ashkin1986observation}, the fields of optomechanics\citep{aspelmeyer2014} and optical trapping have been put into the spotlight of modern research in physics. Recent progress has brought micro-optomechanical systems to the ground state of motion\citep{Oconnell2010, chan2011, Teufel2011}, opening up possibilities for quantum transducers\citep{stannigel2010optomechanical} and force sensors\citep{ranjit2016zeptonewton, monteiro2017optical} while providing new platforms to test quantum mechanics at the mesoscopic scale\citep{bateman2014near, riedinger2018remote, marinkovic2018optomechanical}. 

Such experiments require long coherence times, a property quantified by the mechanical $Q$ factor. To date, the highest $Q$ factors are found in nanoengineered SiN membranes and in levitated particles in vacuum, with values exceeding $10^8$\citep{yuan2015silicon, tsaturyan2017ultracoherent, Gieseler2012}. Being isolated from the environment, levitated particles\citep{yin2013optomechanics} offer further possibilities, since they can be used to study internal phonons, quantized internal degrees of freedom\citep{Rahman2017, conangla2018motion} and matter-wave interferometry. They have been extensively used in previously unaccessible physical regimes\citep{Li2010} and proposed for quantum mechanics experiments\citep{Chang2010cavity, romero2010toward}. 

As with clamped resonators, a general prerequisite of these proposals is the ground state of motion, which so far has remained elusive for levitated systems. Ongoing efforts concentrate on cavity\citep{kiesel2013cavity, mestres2015cooling} and feedback\citep{Li2011, Gieseler2012} cooling of the center of mass (CoM) motion of optically levitated particles, with parametric feedback cooling (PFC)\citep{Gieseler2012} being the current standard technique for motion control and the only to report sub mK temperatures\citep{Jain2016}. An all-electrical feedback approach for highly charged particles has also been proposed\citep{goldwater2018levitated} based on the recent development of charge control in nanoparticles\citep{moore2014search, Frimmer2017Controlling}. However, the separation of feedback force and trapping potential will add flexibility and allow for optimal control (OC) protocols\citep{kwakernaak1972linear}. 

For linear observable systems, the OC law is known as the linear quadratic regulator (LQR) and is widely utilized in larger mechanical systems\citep{kwakernaak1972linear}. It guarantees that a dynamical system will minimize its energy in the fastest way possible. For a levitated nanoparticle, the LQR takes the law of a proportional-derivative controller with optimal gain coefficients. These can be determined analytically, but an additional machine learning (ML) algorithm will autonomously find the optimal gains without prior knowledge of the system parameters.

\begin{figure}
\begin{center}
\includegraphics[width=0.46\textwidth]{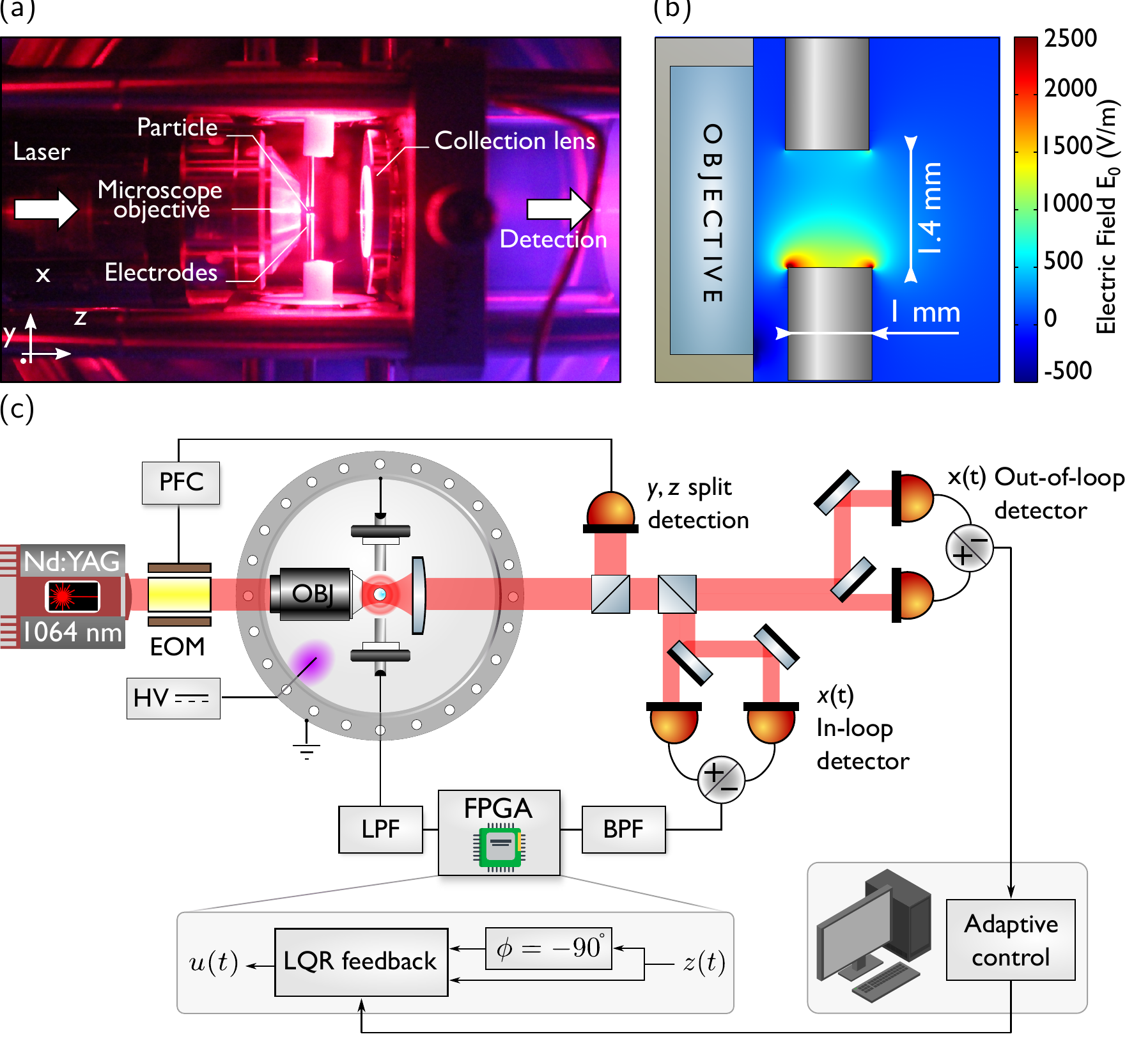}
\caption{\textbf{Experimental setup} (a) Picture of the setup inside the vacuum chamber. The purple glow on the right is a plasma generated to control the particle's charge (b) Close-up image of the trapping region with a simulation of the electric field for an applied DC voltage of $1~{\rm V}$. (c) Setup sketch. A microscope objective (OBJ) is used to trap a silica nanoparticle. The scattered light is collected and sent to a balanced photodiode to detect the motion. The $x$--motion signal is band-pass filtered (BPF) and sent to the FPGA, where the LQR signal is calculated. The output is low-pass filtered (LPF) and sent to the electrodes to cool the motion electrically. The LQR parameters are controlled with a ML routine that runs on the board CPU. Parametric feedback cooling (PFC) is used to modulate the laser light with an electro-optic modulator (EOM) and cool the two other axes during all the measurements.}
\label{fig:1_setup}
\end{center}
\end{figure}

In this letter we present the first demonstration of cooling and control of one mode of the CoM motion of an optically levitated charged nanoparticle with a ML controlled LQR feedback, using electric fields to exert a force on the particle's motion. With a considerably simpler experimental setup than previous feedback experiments\citep{Li2011, Gieseler2012}, the LQR yields temperatures that are between one and two orders of magnitude lower than PFC\citep{Gieseler2012} over the $10^{-1}$ mbar to $10^{-7}$ mbar range and has transients 10 to 600 times faster than regular cold damping. The minimum temperature is eventually limited by the present detection noise floor, yielding a temperature of 5 mK at $3\times 10^{-7}$ mbar.

\vspace{0.5cm}
\textbf{Theory}. The CoM motion along the $x$--axis of an optically levitated particle can be described by the stochastic differential equation
\begin{align}\label{eq:motion}
\ddot{x} + \Gamma \dot{x} + \omega_0^2 x = \frac{\sigma}{m} \eta(t) + u(t),
\end{align}
where $m$ is the particle mass\citep{ricci2018accurate}, $\Gamma$ is the damping term due to the interaction with residual gas molecules, $\omega_0$ is the oscillator's natural frequency, $\sigma \eta(t)$ is a stochastic force with zero mean and autocorrelation $R(\tau) = \sigma^2\delta(\tau)$, associated with the damping via the fluctation-dissipation relation\citep{Kubo1966} $\sigma = \sqrt{2 k_B T m \Gamma}$, and $u(t)$ is an externally applied feedback force of arbitrary form. Experimentally, the velocity $v(t)$ is inaccessible. We can only measure a noisy position, $z(t) = x(t) + \xi(t)$, where $z(t)$ is the \emph{observed} position and $\xi(t)$ is detection noise, in our case dominated by shot noise.

If equation \eqref{eq:motion} describes the system evolution accurately, there exists an OC law\citep{kwakernaak1972linear} $u(t)$ that minimizes the expected energy functional
\begin{align}
\mathcal{J} = \mathbb{E}\left[ \int_0^\mathcal{T} \left(x^2(t) + \rho u^2(x(t))\right) \dif t\right],
\end{align}
where $\rho$ is a weighting parameter and $\mathcal{T}$ is the energy integration time; both can be chosen at will. The expression of $u(t)$ that minimizes $\mathcal{J}$ is given by the LQR controller\citep{kwakernaak1972linear}:
\begin{align}\label{eq:LQR}
u(t) = \mathbf{K} \cdot 
\begin{pmatrix}
x(t) \\
v(t)
\end{pmatrix}, \quad 
\mathbf{K} = 
\begin{pmatrix}
k_ \text{p} & k_ \text{d} \\
\end{pmatrix},
\end{align}
where $\mathbf{K}$ is a constant matrix whose coefficients can be calculated numerically (supplementary).

Since we can make $\rho$ arbitrarily small, then, for a fixed $\mathcal{T}$ the OC law will minimize $\mathcal{J} \propto \langle x^2(t) \rangle$. Therefore, a proportional-derivative controller with expression $u(t) = k_ \text{p} x(t) + k_ \text{d} v(t)$ will minimize the energy among all other possible feedback protocols $u(x(t))$, either linear or nonlinear\citep{kwakernaak1972linear}. In particular, it outperforms PFC\citep{Gieseler2012}, which relies on a modulation of the potential. The case where only a damping term is considered ($k_ \text{p} = 0$) is usually known as cold damping (CD)\citep{bushev2006feedback}. While the final minimal temperature for CD and LQR in most experimental conditions is the same, the LQR has significantly shorter transient times. 

The optimal controller can be separated in two steps: firstly, an optimal phase state estimator (known as Kalman filter\citep{kalman1960, kalman1961, setter2018real}), that will produce estimates of $(x(t), v(t))$ given noisy position measurements $z(t)$; secondly, an optimal feedback (LQR) based on \eqref{eq:LQR}. The combination of both is known as a linear quadratic Gaussian (LQG) controller. In our experiment, instead of a Kalman filter we approximate the phase space coordinates as
$$
(\hat{x}(t), \hat{v}(t)) = (z(t), -\omega_0 z(t - \phi/\omega_0)),
$$
with $\phi = \pi/2$. Not using a Kalman filter yields higher CoM temperatures, but results in a considerably simplified digital signal processing unit. Additionally to the feedback we have implemented a ML algorithm that autonomously optimizes the parameters $(k_ \text{p}, k_ \text{d})$ by minimizing $\langle x^2(t) \rangle$, adapting itself to different experimental conditions.

\begin{figure}[h!]
\begin{center}
\includegraphics[width=0.46\textwidth]{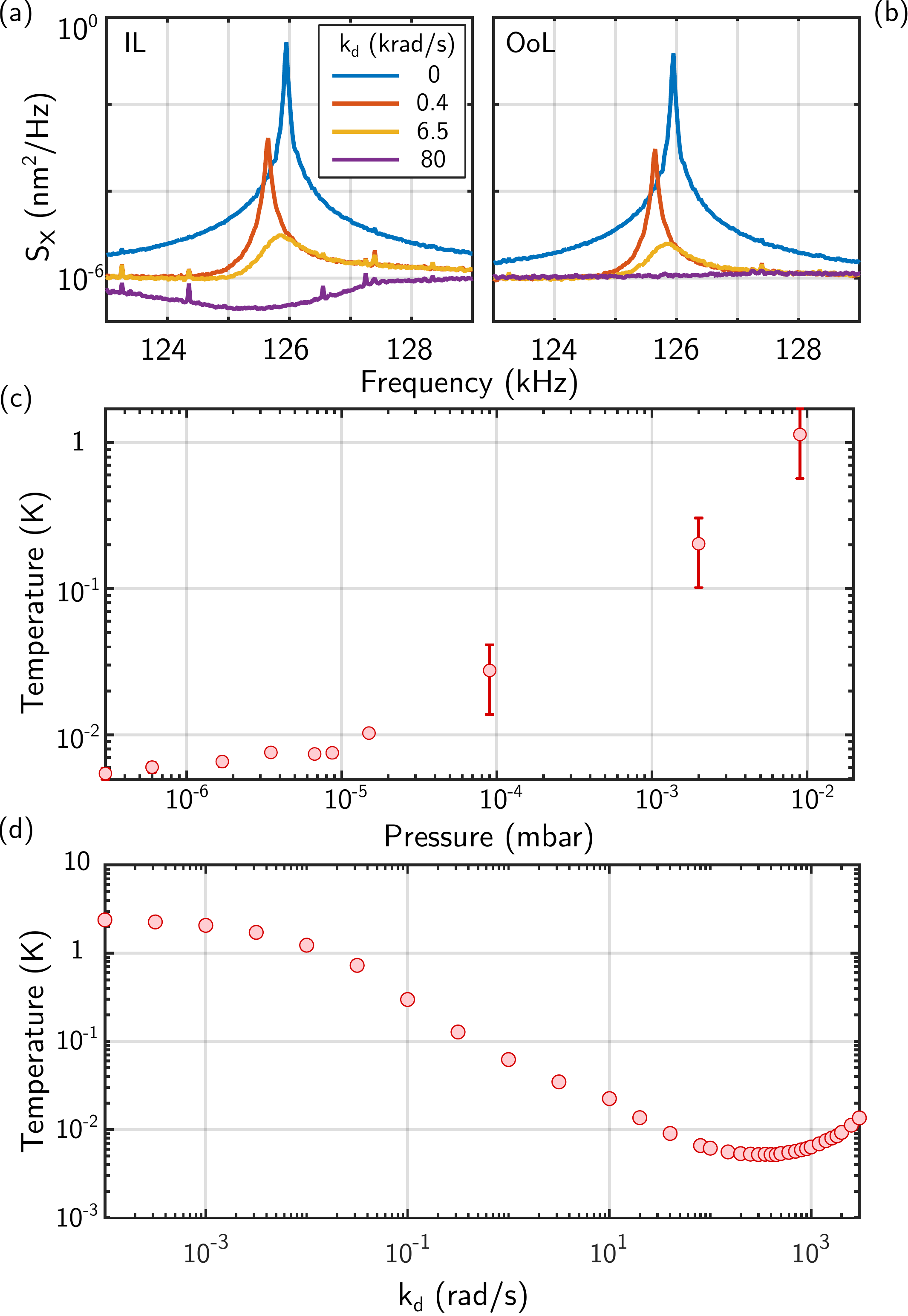}
\caption{\textbf{Cold damping}: $k_ \text{p} = 0$ and varying $k_ \text{d}$. (a) PSD of the motion in the $x$ direction recorded from the IL detector signal at various gains for a particle at $p = 10^{-5}$ mbar. For large $k_ \text{d}$, $S_x(\omega)$ shows noise squashing. (b) PSD of the motion in the $x$ direction simultaneously recorded with the OoL detector. Noise squashing is avoided. (c) Minimum temperature achieved at different pressures for the optimal $k_ \text{d}$ (OoL detector). The lowest temperature, $T_\text{eff} = 5.1 \pm 0.5$ mK, is reached at $3\times 10^{-7}$ mbar. (d) Temperature vs. $k_ \text{d}$ gain at $3 \times 10^{-7}$ mbar, displaying an optimal gain minimum at $k_ \text{d} \simeq 700$ rad/s, where $T_\text{eff} = 5.1 \pm 0.5$ mK (OoL detector). Eq. \eqref{eq:T_mode2} predicts a qualitative behaviour in accordance with data, with optimal $k_ \text{d}$ a factor 3.4 larger than measured. The minimum temperature in the plot is approximately 10 times larger than values predicted by theory.}
\label{fig:2_cold_damping}
\end{center}	
\end{figure}

Since the random thermal noise $\eta(t)$ and measurement noise $\xi(t)$ are uncorrelated, we may calculate (supplementary) the power spectral density (PSD) of the particle position $x(t)$ as
\begin{align}\label{eq:psd}
S_x(\omega) =& \frac{\sigma^2/m^2}{(\omega_0^2 + k_ \text{p} - \omega^2)^2 + (\Gamma + k_ \text{d})^2\omega^2} + \nonumber\\
& \frac{k_ \text{p}^2 + k_ \text{d}^2\omega^2}{(\omega_0^2 + k_ \text{p} - \omega^2)^2 + (\Gamma + k_ \text{d})^2\omega^2}\sigma^2_\xi,
\end{align}
where $S_\xi(\omega) = \sigma^2_\xi$ is the detection noise level (constant at the spectral range of interest in our experiment). The second term of the PSD, absent in freely oscillating particles, is due to the introduction of  noise by the feedback and becomes dominant for large $k_ \text{p}$, $k_ \text{d}$ gains. The feedback also introduces a correlation between detection noise and position that affects the PSD shape of the measured $z(t)$. This $S_\text{IL}(\omega)$, obtained through the in-loop (IL) detector, differs from the expression in eq. \ref{eq:psd} (supplementary).

For small $k_ \text{d}$ the difference between $S_\text{IL}(\omega)$ and $S_x(\omega)$ is negligible. However, due to the correlation of detection noise and $x(t)$, $S_\text{IL}(\omega)$ shows a reduction or \emph{squashing} of the noise floor around $\omega_0$ for large values of $k_ \text{d}$. To avoid underestimations of the particle's effective temperature $T_\text{eff} = m\omega_0^2 \langle x^2\rangle/k_B$ (supplementary), we introduce a second, out-of-loop (OoL), detector with uncorrelated noise. This OoL was omitted in previous levitodynamics feedback cooling experiments\citep{Li2011, Gieseler2012, Jain2016}.

\vspace{0.5cm}
\textbf{Experimental setup}. The experimental setup is displayed in Fig.~\ref{fig:1_setup}. A silica nanoparticle ($235 \pm 11~{\rm nm}$ in diameter) is loaded at ambient pressure into a single beam optical trap inside a vacuum chamber (wavelength $\lambda = 1064~{\rm nm}$, power $P\simeq 75~{\rm mW}$, objective $\text{NA}=0.8$). The charge $Q$ of the particle, $-50$ net e$^+$ in this study, is controlled\citep{Frimmer2017Controlling, ricci2018accurate} with a corona discharge on a bare electrode; the voltage polarity determines the sign of the charges that are added. Along the horizontal direction $x$, a pair of electrodes separated by $d_\text{el}$ form a parallel-plate capacitor around the particle position (Fig.~\ref{fig:1_setup}(b)). Applying a voltage $V(t)$, we create a feedback force $u(t) \approx Q V(t)/d_\text{el}$ on the particle. 

\begin{figure}[h!]
\begin{center}
\includegraphics[width=0.46\textwidth]{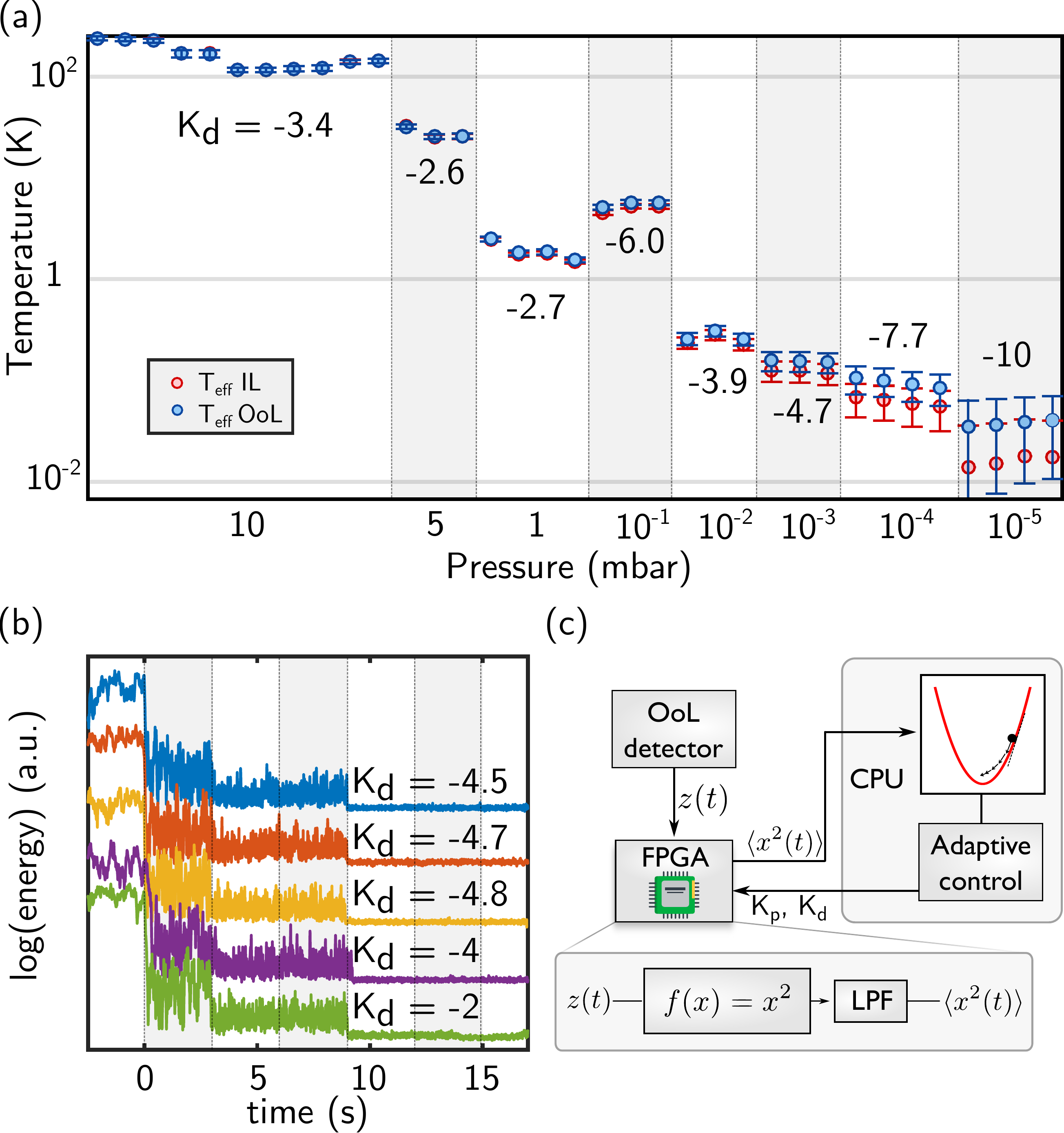}
\caption{\textbf{ML algorithm} (a) ML routine adapting to the optimal $k_ \text{d}$ while the pressure in the chamber is reduced from $p = 10$ mbar to $p = 10^{-5}$ mbar. Points in the same segment are measured at constant pressure. (b) Convergence of the ML routine (started at $t = 0$) to the optimal gain in a few steps at $p = 3\times 10^{-6} $ mbar, for five uncooled $x$ initial conditions. All reach similar final energy values (value of $k_ \text{d}$ in FPGA units is displayed), but slightly differ in the converged gains. Each section represent 3 second iteration steps during which $k_ \text{d}$ has been left constant (for clarity). Data is smoothed with a moving average filter and separated on the vertical axis by a constant factor for readability. (c) Setup and processing details of the algorithm. The OoL signal is sent to the FPGA (through a different input than the IL feedback signal), where the mode energy is estimated. A stochastic gradient descent algorithm, running on the CPU, optimizes the values of $k_ \text{p}$, $k_ \text{d}$.}
\label{fig:3_ML}
\end{center}	
\end{figure}

Figure \ref{fig:1_setup}(c) shows a sketch of the optical setup. We use balanced photodiodes to monitor the oscillation of the particle over all three degrees of freedom. Along the two oscillation modes perpendicular to $x$ we perform PFC\cite{Gieseler2012}. This maintains the particle motion in these two directions in the linear regime, avoiding mechanical cross coupling with the $x$ mode, while keeping the particle trapped at high vacuum. The CoM position in the $x$ direction is detected with two photodiodes: the first, an IL detector, generates the feedback signal used to cool the particle with the LQR, whereas the OoL detector solely records data.

The IL $x$ signal is first processed with an analog band-pass filter, then sent to a FPGA where it is separated into $\hat{x}(t)$ and $\hat{v}(t)$ by delaying the signal appropriately, amplified with the $k_ \text{p}$, $k_ \text{d}$ gains and summed back together. The resulting $u(t)$ feedback signal is later low-pass filtered and sent to the electrodes (see Fig.\ref{fig:1_setup}(c)). The feedback gains $k_ \text{p}$ and $k_ \text{d}$ are controlled from the board CPU, either manually or with a ML algorithm that adapts the gains autonomously. This adaptive routine obtains an estimate of the particle energy from the OoL detector signal and finds the gain values that minimize it with a stochastic gradient descent technique\citep{friedman2001elements}.

\vspace{0.5cm}
\textbf{Results} Cold damping measurements ($k_ \text{p}, k_ \text{d}$ manually set) at pressures ranging from $10^{-2}$ mbar to $10^{-7}$ mbar are shown in Fig. \ref{fig:2_cold_damping}. Figures \ref{fig:2_cold_damping}(a) and \ref{fig:2_cold_damping}(b) display the double sided PSDs $S_x^{\text{IL}}(\omega)$ and $S_x^\text{OoL}(\omega)$, measured from the IL and OoL detectors respectively. For strong feedback gains in the order of $k_ \text{d} = 1$ krad/s the energy of the mode becomes comparable to the noise energy. In contrast to the OoL measurement, on the IL PSDs we observe squashing of the noise floor due to the correlation of detector noise and particle signal. Noise squashing was previously observed in other systems\citep{cohadon1999cooling, Poggio2007, wilson2015measurement} but never in feedback cooled levitated nanoparticles. In Fig. \ref{fig:2_cold_damping}(c) we plot the minimum CoM temperature for different pressures, achieving temperatures between one and two orders of magnitude lower than PFC\citep{Gieseler2012}, since PFC is nonlinear and becomes inefficient for small $x(t)$. Below $p < 10^{-5}$ mbar $T_\text{eff}$ is not reduced as efficiently as at higher pressures, suggesting the detectors noise floor $\sigma_\xi^2$ as the main limiting factor at this pressure range. In Fig. \ref{fig:2_cold_damping}(d) we show $T_\text{eff}(k_ \text{d})$ at $3 \times 10^{-7}$ mbar, achieving a minimum $T_\text{eff} = 5.1 \pm 0.5$ mK. The qualitative temperature behaviour agrees with theory, although the expected optimal $k_ \text{d}$ is a factor 3.4 larger than measured and the minimum $T_\text{eff}$ is approximately 10 times larger than predicted with the theoretical expression of $T_\text{eff}$ (supplementary). Since at these temperatures the motion PSDs are close to the noise floor, this could be due to noise correlations or PFC effects unaccounted for in the model. 
 
In Fig. \ref{fig:3_ML} we present time traces with the ML adaptive algorithm on. Figure \ref{fig:3_ML}(a) displays the particle's temperature as pressure is reduced, showing how the algorithm adapts $k_ \text{d}$ to different conditions starting from a hot particle (individual data points are recorded at constant pressure while the algorithm is in standby). At $10^{-5}$ mbar, temperatures lie in the $10^{-2}$ mK range, recovering the results obtained with the cold damping pressure scan. Figure \ref{fig:3_ML}(b) shows how the algorithm converges to similar final temperatures for five identical initial conditions. Fluctuations of the estimated energy result in a ``noisy'' convergence of the algorithm around the optimum, a characteristic feature of stochastic gradient descent based techniques. A block diagram of the algorithm processing details is sketched in Fig. \ref{fig:3_ML}(c) (details in supplementary).

\begin{figure}[h!]
\begin{center}
\includegraphics[width=0.46\textwidth]{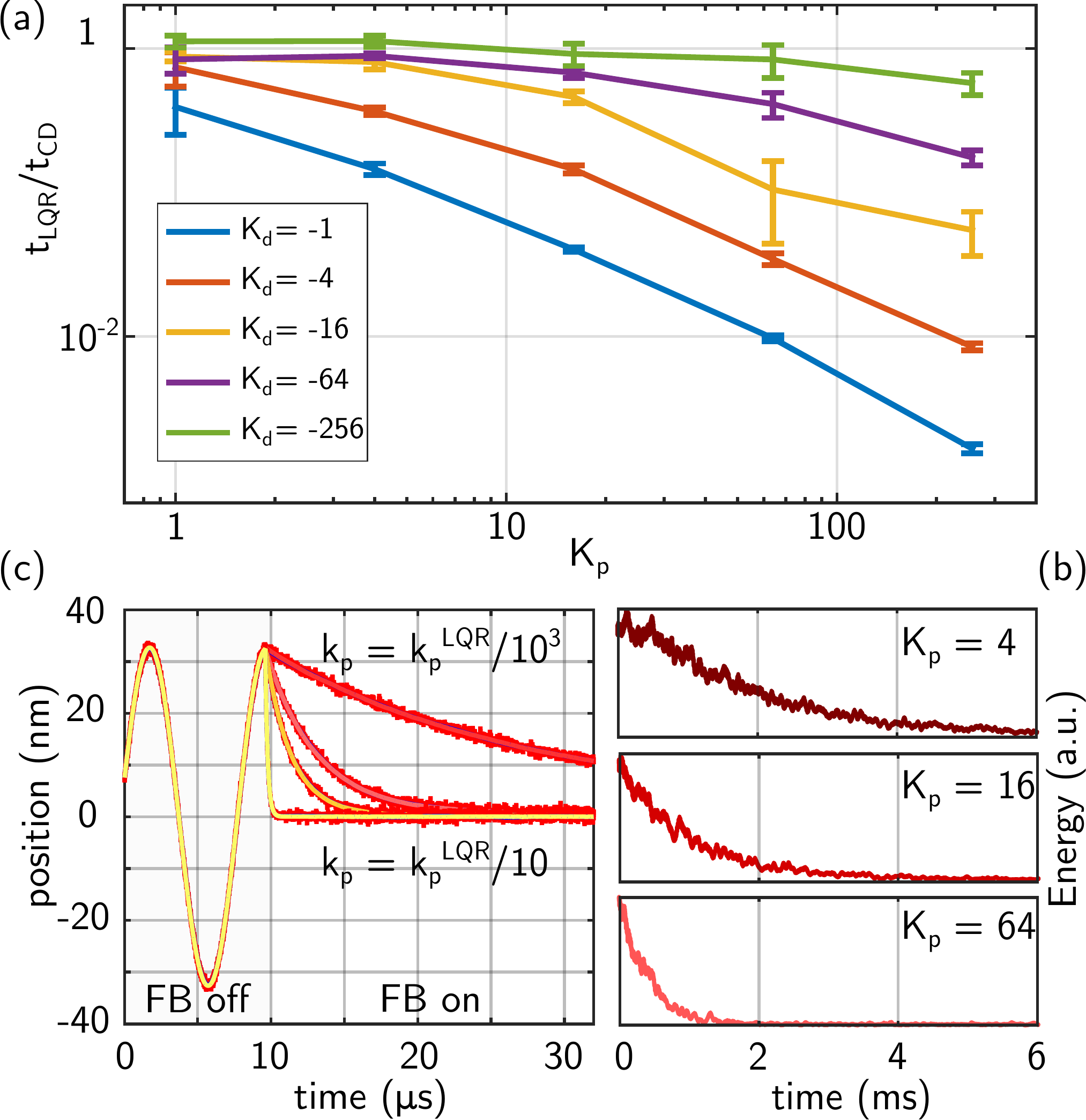}
\caption{\textbf{LQR} (a) Ratio $t_\text{LQR}/t_\text{CD}$ of the decay times (energy reduced to $1/e$) of a controller having both $K_ \text{p}$ and $K_ \text{d}$ (gain values in arbitrary FPGA units) and of cold damping ($k_ \text{p} = 0$) for a pressure $p = 10^{-5}$ mbar. The figure displays how the transient times can be reduced by adding a $K_ \text{p}$ term to the feedback controller. Lines connecting points are added as guides to the eye. (b) Experimental sample traces of the transients after turning on the feedback ($t = 0$) for a fixed value of $k_ \text{d}$. The transients are reduced by increasing $k_ \text{p}$. (c) Numerical simulation of controllers approaching the $k_ \text{p}^\text{LQR}$ optimal value. From red (smaller $k_ \text{p}$) to yellow (larger and closer to $k_ \text{p}^\text{LQR}$). The simulations show how the minimum stationary energy can be reached in a fraction of one oscillation period when $k_ \text{p}$ is properly tuned. In our measurements we never reach $\mu$s transients, due to the use of artificial delays to approximate $v(t)$.}
\label{fig:4_LQR}
\end{center}	
\end{figure}

Finally, introducing a $k_ \text{p}$ term, we investigate (Fig. \ref{fig:4_LQR}) the full LQR and the difference in transient times in comparison to cold damping; we also compare the results with full system simulations of the LQG. An increase in $k_ \text{p}$ leads to faster cooling, as shown in Fig. \ref{fig:4_LQR}(a). Here, the ratio of decay times between cold damping $t_\text{CD}$ and OC's $t_\text{LQR}$ with increasing values of $k_ \text{p}$ is plotted for different values of $k_ \text{d}$, showing decay times a factor 10 to 600 shorter. Three experimental sample paths serve as examples in Fig. \ref{fig:4_LQR}(b). Reducing the time required to transition between different thermal states is beneficial in experiments where the feedback signal needs to adapt quickly to avoid particle loss or where high repetition rates are required. However, as shown in Fig. \ref{fig:4_LQR}(c), the experimental transient times that we observe are still orders of magnitude longer than the simulated LQG. This is probably due to the introduction of artificial delays to approximate $v(t)$, which reduce the correlation between feedback signal and actual phase space variables.

\vspace{0.5cm}
\textbf{Conclusions and Outlook}. In summary, we have demonstrated a novel feedback cooling technique for levitated nanoparticles based on an adaptive optimal protocol, using electric fields to act on charged particles. Our feedback scheme, which only requires an FPGA and electrodes in the vicinity of the levitated particle, stands out for its robustness and simplicity, and can be easily extended to 3D cooling. Importantly, the usage of OoL detection addresses the limitations of prior implementations\citep{Li2011, Jain2016}, preventing potential energy underestimations and hence providing a more accurate temperature readout. In the present experiment we reach a temperature of 5 mK at $3\times 10^{-7}$ mbar, corresponding to an occupation number below 1000 phonons. 

Shot noise and the optical setup measurement efficiency set the lowest achievable temperature; since detection was not optimized, this leaves potential for future improvements. Detection efficiency can be greatly increased by exploiting cavity enhanced detection schemes. If additionally the experiment is performed at ultra-high vacuum, occupation numbers can be brought down by a factor $10^2$ or $10^3$, making ground-state cooling attainable\citep{rossi2018measurement}.

In our scheme, the transients between thermal and cooled states are at least one order of magnitude shorter than in regular cold damping. This feature may be important in experiments where sudden changes (such as varying optical potentials when approaching a surface or nanostructure) might lead to particle loss. Furthermore, the ML algorithm optimizes the cooling performance continuously, adapting to different regimes without requiring prior knowledge of the initial particle state within $\tau_\textsubscript{ML} < 5$ s. This makes it especially suitable for experiments with slowly varying conditions, such as pressure or intensity, minimizing the need for continuous realignment and feedback optimization. Future extension to LQG control will reduce the minimum achievable temperature two-fold, since the use of a Kalman filter will produce optimal estimates of $v(t)$, reducing the effect of measurement noise, and eliminate the need for artificial delays in the system. The introduction of further parameters from the Kalman filter will make the ML algorithm indispensable, since optimization will become a high dimensional problem.

We anticipate that the presented adaptive feedback technique can be implemented in a diverse range of levitodynamics experiments, since it lends easily to miniaturization and automation. It can be a significant addition in studies requiring robustness and high repetition rates, like the planned future space mission MAQRO\citep{kaltenbaek2012macroscopic}, or in small devices, such as force and inertial sensors\citep{barbour2001inertial} based on levitated objects. 

\vspace{0.5cm}
\emph{Note added}. We have recently become aware of related work performed by Tebbenjohanns et al. at ETH Zurich and Iwasaki et al. at the Tokyo Institute of Technology.

\vspace{0.5cm}
\textbf{Authors' contributions}.\hspace{0.2cm}G.P.C. conceived the idea, programmed the FPGA and performed numerical calculations. F.R. developed the optical setup. G.P.C., M.T.C. and F.R. performed the measurements. G.P.C. and N.M. did analytical calculations. N.M., G.P.C. and F.R. processed the experimental data. A.W.S. contributed to the feedback idea. R.Q. supervised the project. G.P.C and N.M. wrote the manuscript with input from all authors.

\vspace{0.5cm}
\textbf{Acknowledgments}.\hspace{0.2cm}The authors acknowledge financial support from the European Research Council through grant QnanoMECA (CoG - 64790), Fundació Privada Cellex, CERCA Programme / Generalitat de Catalunya, and the Spanish Ministry of Economy and Competitiveness through the Severo Ochoa Programme for Centres of Excellence in R$\&$D (SEV-2015-0522), grant FIS2016-80293-R. 

G.P. Conangla thanks J. Martínez and S. López for their help with FPGA programming. The authors thank I. Alda for her help revising the text.

\bibliographystyle{apsrev4-1}
\bibliography{./references}

\newpage
\section*{Supplementary Material}

\subsection*{Code and data}

The FPGA (Red Pitaya STEMlab board) bitstream has been programmed in Vivado Design Suite, combining Xilinx IP cores and custom Verilog code. The feedback law (gains, delay, machine learning on/off, etc.) is controlled from the Red Pitaya CPU board with custom made C code that communicates with the FPGA through registers. Links to the code (which can be downloaded and freely used) can be found \href{www.levitodynamics.icfo.eu}{here}.

Data from measurements and MATLAB code used for the analysis can also be found \href{www.levitodynamics.icfo.eu}{here}.

\subsection*{Spectral densities and $T_\text{eff}$}

The CoM motion along the $x$--axis of an optically levitated particle subject to a LQR (which takes the expression of a proportional-derivative feedback controller in this case) is described by the stochastic differential equation
\begin{align}\label{eq:sup_motion}
\ddot{x} + \Gamma \dot{x} + \omega_0^2 x &= \frac{\sigma}{m} \eta(t) -k_ \text{p}(x + \xi(t))\nonumber \\
& -k_ \text{d}(v(t) + \dot{\xi}(t)),
\end{align}

where $m$ is the particle mass, $\Gamma$ is the damping term due to the interaction with air molecules, $\omega_0$ is the oscillator natural frequency, $\sigma \eta(t)$ is a stochastic force with zero mean and autocorrelation $R(\tau) = \sigma^2\delta(\tau)$, associated with the damping via the fluctation-dissipation relation\citep{Kubo1966} $\sigma = \sqrt{2 k_B T m \Gamma}$, $k_ \text{p}$ and $k_ \text{d}$ are the feedback gains and $\xi(t)$ is a signal representing measurement noise. The model described by eq. \eqref{eq:sup_motion} is accurate as long as the optical field is well approximated by a quadratic potential. This is usually the case at pressures above 50 mbar, where the viscous damping dominates the particle's dynamics, and is also a good description at lower pressures when the feedback $u(t)$ restricts the particle's motion to the vicinity of the optical trap center. 

Taking the Fourier transform of \eqref{eq:sup_motion}, defining $\mathcal{F}(\xi(t)) = \Lambda_\xi(\omega)$, $\mathcal{F}(\eta(t)) = \Lambda_\eta(\omega)$ and solving for $X(\omega)$ we get
\begin{align}
X(\omega) & = \frac{\sigma/m}{(\omega_0^2 + k_ \text{p} - \omega^2) + i\omega(\Gamma + k_ \text{d})}\Lambda_\eta(\omega) \nonumber \\
&- \frac{(k_ \text{p} + ik_ \text{d}\omega)\Lambda_\xi(\omega)}{(\omega_0^2 + k_ \text{p} - \omega^2) + i\omega(\Gamma + k_ \text{d})}.
\end{align}

Equivalently, the Fourier transform including the measurement noise will be
\begin{align}
X(\omega) + \Lambda_\xi(\omega) = \frac{\sigma/m}{(\omega_0^2 + k_ \text{p} - \omega^2) + i\omega(\Gamma + k_ \text{d})}\Lambda_\eta(\omega) - \nonumber\\
 \frac{(w_0^2-\omega^2) + i\omega\Gamma}{(\omega_0^2 + k_ \text{p} - \omega^2) + i\omega(\Gamma + k_ \text{d})}\Lambda_\xi(\omega). 
\end{align}

By using $S_x(\omega) = \mathbb{E}(|X(\omega)|^2)$ and the fact that $\xi(t)$ and $\eta(t)$ are uncorrelated we find the PSDs of both the real position and the measured position in the IL detector:
\begin{align}
S_x(\omega) &= \frac{\sigma^2/m^2}{(\omega_0^2 + k_ \text{p} - \omega^2)^2 + (\Gamma + k_ \text{d})^2\omega^2} + \nonumber\\
 &\frac{k_ \text{p}^2 + k_ \text{d}^2\omega^2}{(\omega_0^2 + k_ \text{p} - \omega^2)^2 + (\Gamma + k_ \text{d})^2\omega^2}\sigma_\xi^2,\\
S_\text{IL}(\omega) &= \frac{\sigma^2/m^2}{(\omega_0^2 + k_ \text{p} - \omega^2)^2 + (\Gamma + k_ \text{d})^2\omega^2} + \nonumber\\
&\frac{(\omega_0^2 - \omega^2)^2 + \Gamma^2\omega^2}{(\omega_0^2 + k_ \text{p} - \omega^2)^2 + (\Gamma + k_ \text{d})^2\omega^2}\sigma_\xi^2.\label{eq:psds}
\end{align}

With a completely analogous argument, if we define $\beta(t)$ as the measurement noise in the OoL detector, the measured OoL position PSD will be
\begin{align}
S_\text{OoL}(\omega) &= \frac{\sigma^2/m^2}{(\omega_0^2 + k_ \text{p} - \omega^2)^2 + (\Gamma + k_ \text{d})^2\omega^2} + \nonumber\\
& \frac{k_ \text{p}^2 + k_ \text{d}^2\omega^2}{(\omega_0^2 + k_ \text{p} - \omega^2)^2 + (\Gamma + k_ \text{d})^2\omega^2}\sigma_\xi^2 + \sigma_\beta^2.
\end{align}

\begin{figure}[h!]
\begin{center}
	\includegraphics[width=0.45\textwidth]{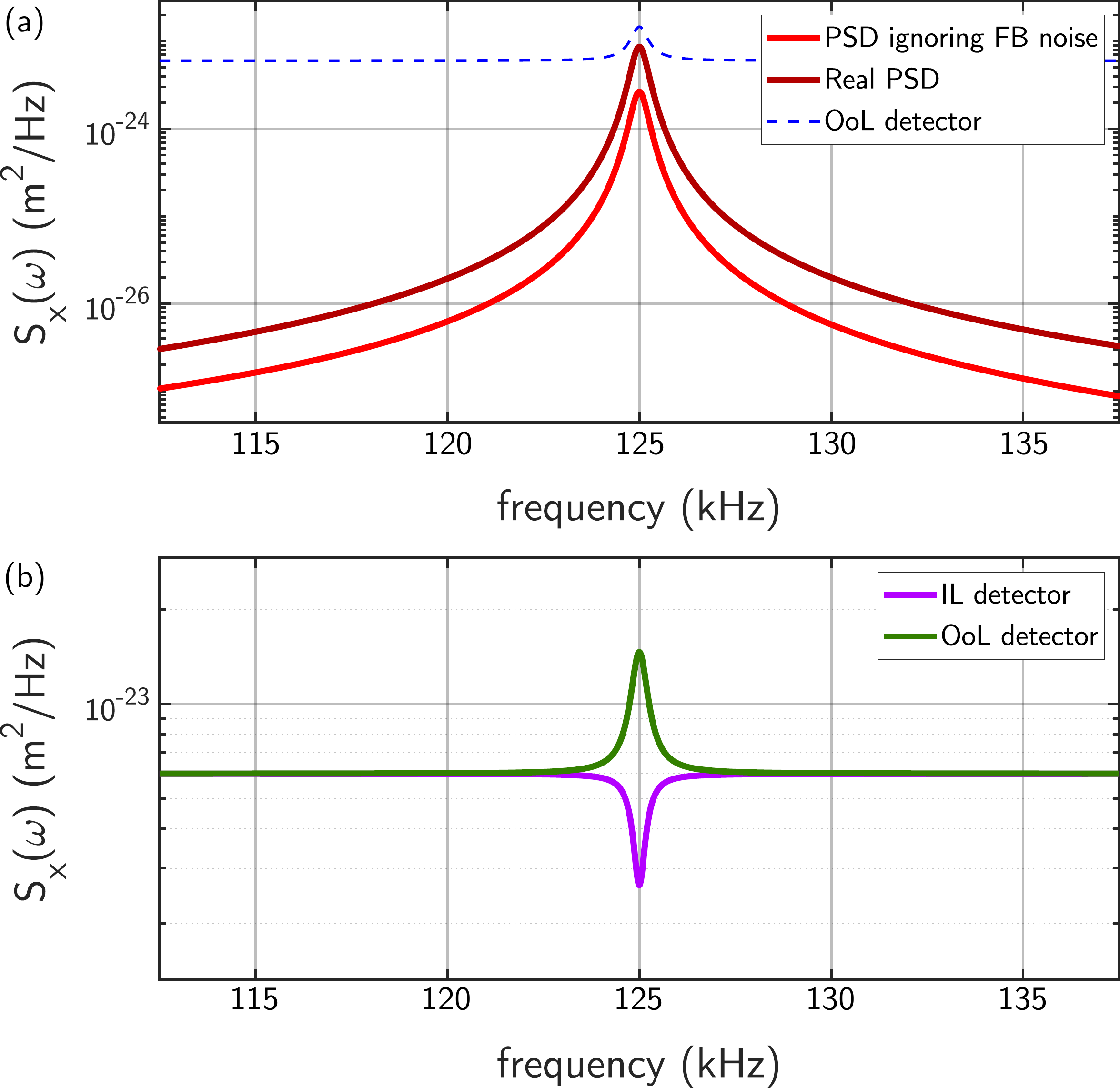}
	\caption{Theoretical PSD of the CoM motion in the presence of cold damping with a perfectly adjusted delay $\phi = \pi/2$. (a) The plot shows three PSDs: the particle's PSD without taking into account measurement noise, the real $S'_x(\omega)$ and the signal $S'_\text{OoL}(\omega)$ measured in the out of loop. The PSD $S_x(\omega)$ would practically overlap with $S'_x(\omega)$. (b) Comparison of the PSD signals measured in the IL detector and OoL detector, showing perfect squashing symmetry in the IL.}
	\label{fig:s2_psds}
\end{center}
\end{figure}

In this experiment, however, we haven't used the real particle velocity. Instead, we have approximated 
$$
v(t) \simeq \omega_0 x(t - \tau),
$$
where $\tau = \phi/\omega_0$ and $\phi = \pi/2$. By using the fact that $\mathcal{F}(x(t - \tau)) = e^{-i\omega \tau}X(\omega)$ we may obtain new expressions for the PSDs. Considering only a derivative gain $k_ \text{d}$ and defining
$$
G(\omega)= (\omega_0^2 - k_ \text{d}\cos(\omega\tau) - \omega^2)^2 + (\Gamma\omega + \omega_0k_ \text{d}\sin(\omega\tau))^2,
$$
then the new PSD expressions will be
\begin{small}\begin{align}
S'_x(\omega) & = \frac{\sigma^2/m^2}{G(\omega)} + \frac{\omega_0^2k_ \text{d}^2}{G(\omega)}\sigma_\xi^2,\\
S'_\text{IL}(\omega)&  = \frac{\sigma^2/m^2}{G(\omega)} + \frac{(\omega_0^2 - \omega^2)^2 + \Gamma^2\omega^2}{G(\omega)}\sigma_\xi^2\\
S'_\text{OoL}(\omega) & = \frac{\sigma^2/m^2}{G(\omega)} + \frac{\omega_0^2k_ \text{d}^2}{G(\omega)}\sigma_\xi^2 + \sigma_\beta^2.
\label{eq:psds_delay}
\end{align}\end{small}

\begin{figure}[h!]
\begin{center}
\includegraphics[width=0.45\textwidth]{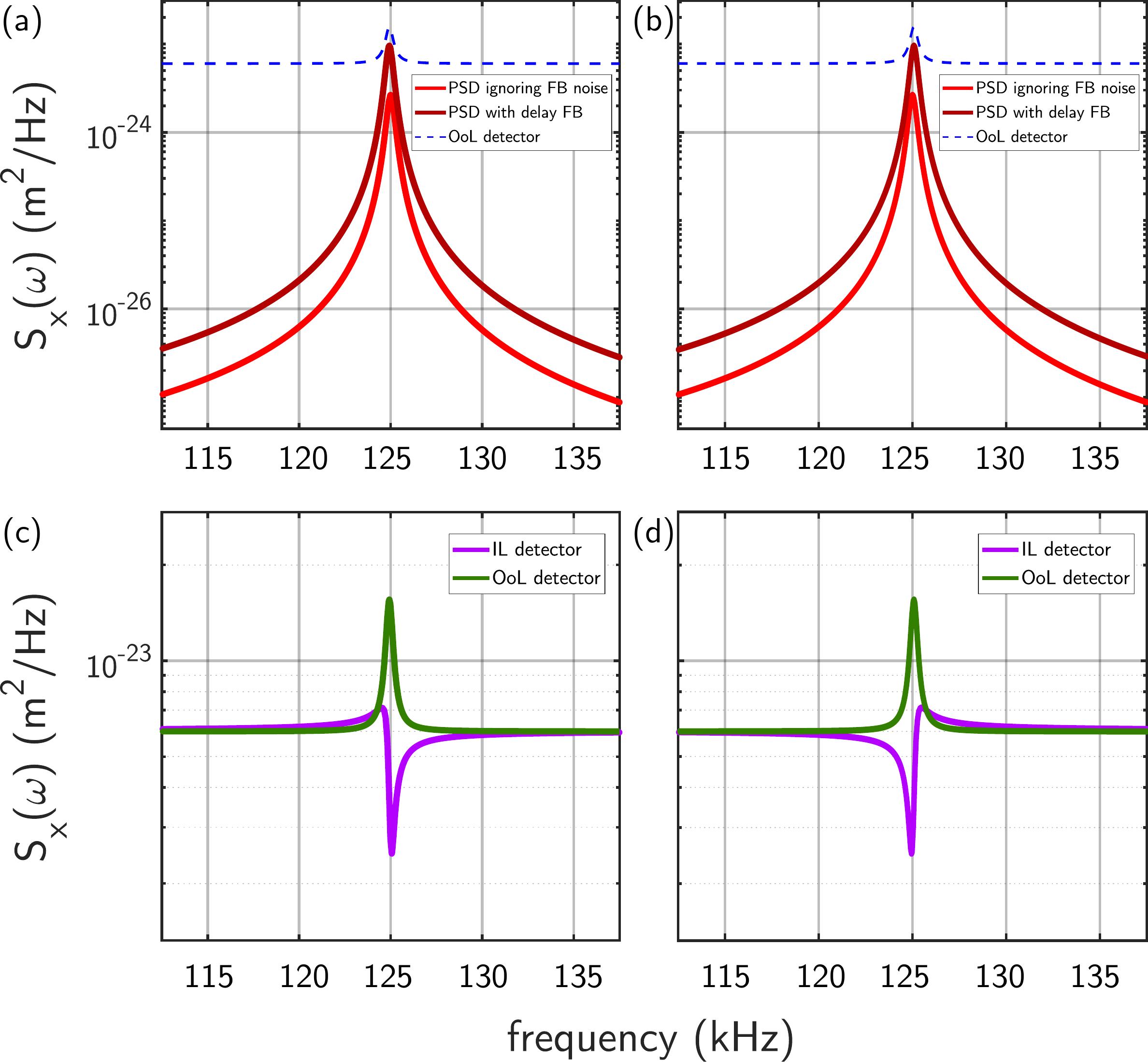}
\caption{Theoretical PSD of the CoM motion in the presence of cold damping with a smaller/larger than $\pi/2$ delay. (a) The plot shows three PSDs: the particle's PSD without taking into account measurement noise, the real $S'_x(\omega)$ and the signal $S'_\text{OoL}(\omega)$ measured in the out of loop, for a delay of $0.8\cdot \pi/2$. (b) The plot shows three PSDs: the particle's PSD without taking into account measurement noise, the real $S'_x(\omega)$ and the signal $S'_\text{OoL}(\omega)$ measured in the out of loop, for a delay of $1.2\cdot \pi/2$. (c) Comparison of the PSD signals measured in the IL detector and OoL detector, showing asymmetry in the IL, for a delay of $0.8\cdot \pi/2$. (d) Comparison of the PSD signals measured in the IL detector and OoL detector, showing asymmetry in the IL, for a delay of $1.2\cdot \pi/2$.}
\label{fig:s1_psds}
\end{center}
\end{figure}

\begin{figure}[h!]
\begin{center}
\includegraphics[width=0.45\textwidth]{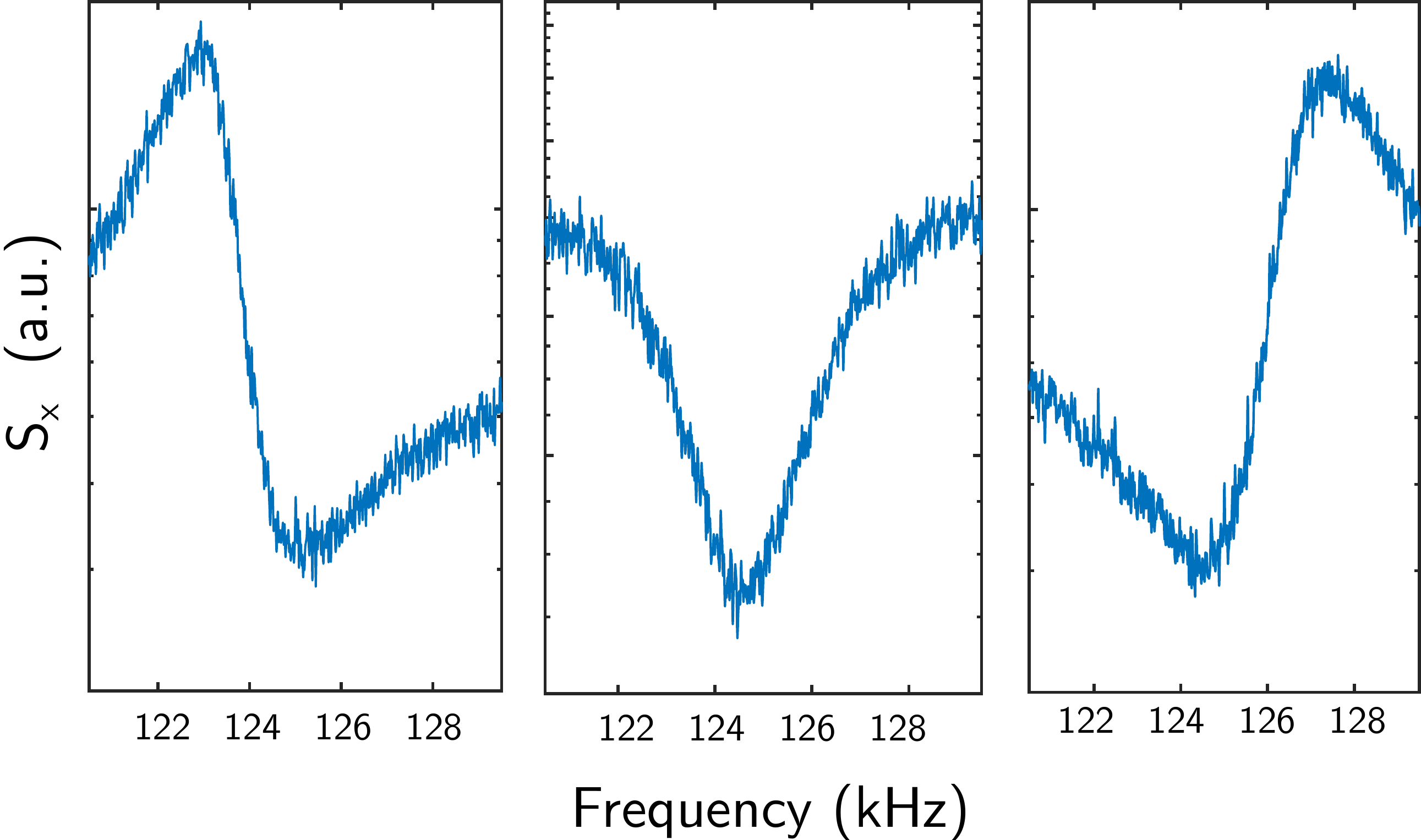}
\caption{Experimental PSDs recorded with the IL detector for large $k_ \text{d}$ gains ($k_ \text{p} = 0$), showing squashing of the noise floor. The center PSD displays a symmetrical squashing dip, corresponding to a feedback delay that is correctly tuned. Decreasing (left PSD) or increasing (right PSD) the value of the delay leads to squashing asymmetry, as predicted in equations \ref{eq:psds_delay} and plotted in Fig. \ref{fig:s1_psds}.}
\label{fig:s3_psds}
\end{center}
\end{figure}

These PSDs are very similar to the ones found before for values of $\omega$ close to $\omega_0$ as long as $\phi$ is exactly $\pi/2$. Nevertheless, for values smaller or larger than $\pi/2$ the resulting PSD will have a small asymmetry, very visible in the IL noise squashing. Figure \ref{fig:s2_psds} shows the resulting theoretical PSDs with a properly tuned delay, whereas the case of $\phi \neq \pi/2$ is displayed in Fig. \ref{fig:s1_psds}. Experimental data with different delay values is shown in Fig. \ref{fig:s3_psds}, showing good agreement with the derived expressions.

Using the equipartition theorem, we define the mode effective temperature as $T_{\text{eff}} = m\omega_0^2\langle x^2 \rangle/k_B$, which we can find with Parseval's theorem as 
$$
T_{\text{eff}} = \frac{m\omega_0^2}{k_B} \frac{1}{2\pi}\int_{-\infty}^\infty S_x(\omega) \dif \omega.
$$
Using the following integral expressions
\begin{align*}
\frac{1}{2\pi}\int_{-\infty}^\infty \frac{1}{(\omega^2-\omega_0^2)^2 + \Gamma^2\omega^2} \dif \omega &= \frac{1}{2\omega_0^2\Gamma}, \\
\frac{1}{2\pi}\int_{-\infty}^\infty \frac{\omega^2}{(\omega^2-\omega_0^2)^2 + \Gamma^2\omega^2} \dif \omega &= \frac{1}{2\Gamma},
\end{align*}
we find
\begin{align}\label{eq:T_mode2}
T_{\text{eff}} = \frac{m\omega_0^2}{2k_B}\left(\frac{\sigma^2/m^2 + k_ \text{p}^2 \sigma_\xi^2}{(\omega_0^2 + k_ \text{p})(\Gamma + k_ \text{d})} + \frac{k_ \text{d}^2 \sigma_\xi^2}{(\Gamma + k_ \text{d})} \right).
\end{align}
which coincides with the expression for cold damping found in \citep{Poggio2007} when $k_ \text{p} = 0$. We use expression \eqref{eq:T_mode2} to compare the measured temperatures with theoretical values in terms of $k_ \text{p}$, $k_ \text{d}$.

\subsection*{Data evaluation details}
We estimate the conversion factor between the FPGA software gain $K_ \text{p}$, $K_ \text{d}$ (in arbitrary units) and $k_ \text{p}$, $k_ \text{d}$ as defined in the equation of motion to be
\begin{align}
k_ \text{p} &\simeq 55 \frac{\text{bits}}{\text{nm}} \frac{10}{32768}\frac{\text{V}}{\text{bits}} G_\text{el} 577 \frac{ne^-}{m}\nonumber\\
& 10^9 \frac{\text{nm}}{\text{m}} K_ \text{p} = 3.34 \times 10^{8} K_ \text{p} \\
K_ \text{d} &\simeq \frac{3.34 10^{8}}{\omega_0} k_ \text{d} = 4.26 \times 10^2 K_ \text{d}  \label{eq:kdKd}
\end{align}

where $G_\text{el}$ accounts for the electronic gain in our setup, $m \approx 2\times 10^{-17}$ kg is the mass of the particle used throughout the letter, $e^-$ the electron charge and $n = 50$ the number of elementary charges in our particle. 

To estimate the energy of the particle's $x$ mode the OoL PSD is background corrected by subtracting the detection noise floor and then the uncalibrated area of the PSD is summed up over a region of interest of approx $\pm 15$ kHz. The calibration (i.e., volts to m$^2$/Hz conversion factor) is obtained by calculating the area of a PSD at 50 mbar without any feedback, assuming that it is thermalized at a room temperature of $T= 295$ K (nonlinear terms of the optical field expansion have contributions way below experimental error at this pressure). Dividing the respective areas and multiplying with the room temperature yields the effective mode temperature.

The uncertainty in the evaluation of temperature is calculated by taking into account the  uncertainties of the PSD, noise floor and calibration factor. They are displayed as error bars in the figure plots.

\subsection*{Machine learning algorithm}

We use a form of stochastic gradient descent\citep{friedman2001elements} for the adaptive feedback algorithm. The objective function to be minimized is an estimation of the particle energy, calculated as
\begin{align}
Q(K_ \text{p}, K_ \text{d}) = h*\left(\zeta^2(t)\right) \approx \langle x^2(t) \rangle,
\end{align}

where $K_ \text{p}$, $K_ \text{d}$ are in arbitrary FPGA units, $\zeta(t)$ is the signal measured in the OoL detector after an analog band-pass filter, and $h[n]$ is a low-pass digital infinite impulse response filter of order 1. This low-pass filter is implemented in the FPGA and has a cutoff frequency $f_\text{c} = 0.3$ Hz, designed to eliminate fluctuations and, thus, the time dependency on $Q$ (not to be confused with the particle's charge).

Experimentally, the OoL signal is fed into a second FPGA input, and $Q$ is continuously calculated. The updated values of $Q$ are written into a register at 62.5 MHz, and custom-made software designed to control the FPGA (running on the board CPU) reads the current energy value. The program decides the new values of $K_ \text{p}$, $K_ \text{d}$ according to
\begin{align}
(K_ \text{p}[n], K_ \text{d}[n]) = (K_ \text{p}[n-1], K_ \text{d}[n-1]) - \delta \nabla Q,
\end{align}

where the step size $\delta$ has been chosen to ensure convergence and reasonable speeds and the gradient of $Q$ is approximated as
\begin{align}
 &\nabla Q(K_ \text{p}[n], K_ \text{d}[n])
\approx \nonumber \\
 &\begin{pmatrix}
\frac{Q(2K_ \text{p}[n], K_ \text{d}[n]) - Q(K_ \text{p}[n], K_ \text{d}[n])}{K_ \text{p}[n]} \\
\frac{Q(K_ \text{p}[n], 2K_ \text{d}[n]) - Q(K_ \text{p}[n], K_ \text{d}[n])}{K_ \text{d}[n]}
\end{pmatrix}
\end{align}
by exploring for a short time different values of $K_ \text{p}$ and $K_ \text{d}$.

\subsection*{Simulations of the LQG}

The LQG minimizes the functional
\begin{align}
\mathcal{J} = \mathbb{E}\left[ \int_0^T \left(x^2(t) + \rho u^2(x(t))\right) \dif t\right],
\end{align}
but since we can make $\rho$ arbitrarily small, then, for a fixed $T$,
\begin{align}
\mathcal{J} \approx T \, \mathbb{E}\left[ \frac{1}{T}\int_0^T x^2(t) \dif t\right] \propto \langle x^2(t) \rangle.
\end{align}
In other words, for $\rho \ll 1$ the LQG minimizes the energy functional among all other feedback laws (a similar argument can be made for $n$--dimensional linear systems). Since we have used some approximations (i.e., no Kalman filter) in the actual implementation of the feedback scheme, we use simulations of the LQG as a benchmark for comparison.

The simulations of the LQG have been performed in MATLAB and consist of a three step process:
\begin{enumerate}
\item We generate a signal $x(t)$ to emulate the particle's position. After that we \emph{generate} and add a measurement noise, obtaining $z(t)$, thus taking into account the two dominant noise sources (shot noise and electronic noise) in the measured signal.
\item The signal $\hat{x}(t)$ is reconstructed from $z(t)$ by a Kalman filter.
\item We add a feedback step with a LQR. We first calculate $k_ \text{p}$, $k_ \text{d}$ by solving the Ricatti equation, as described in \citep{chow1975analysis}, and calculate $u(t)$ as in equation \ref{eq:LQR}. We add $u(t)$ to the equation of motion when the simulated feedback is ``turned on''.
\end{enumerate}

Finally, we compare the results of the LQR with the ones where different values of $k_ \text{p}$ and $k_ \text{d}$ are used. 

The signal $x(t)$ simulation is performed with a Runge-Kutta method of strong order 1\citep{Rossler2009Second}, which we detail in what follows: let $\mathbf{X}(t) \in \mathbb{R}^n$ be the stochastic process that we want to simulate, satisfying the general It\^o stochastic differential equation (SDE):
$$
\dif \textbf{X} = \textbf{a}(t, \textbf{X})\,\dif t+ \textbf{b}(t, \textbf{X})\,\dif W.
$$
Given a time step $\Delta t$ and the value $\textbf{X}(t_k)= \textbf{X}_k$, then $\textbf{X}(t_{k+1})$ is calculated recursively as
\begin{small}
$$
\begin{array}{rl}
\textbf{K}_1 = & \textbf{a}(t_k, \textbf{X}_k) \Delta t  +(\Delta W_k-S_k\sqrt{\Delta t})\cdot \textbf{b}(t_k, \textbf{X}_k),
\\
\textbf{K}_2 = & \textbf{a}(t_{k+1}, \textbf{X}_k+ \textbf{K}_1) \Delta t \,+\\
& (\Delta W_k+S_k\sqrt{\Delta t}) \cdot \textbf{b}(t_{k+1}, \textbf{X}_k+\textbf{K}_1),\\
\textbf{X}_{k+1} = & \textbf{X}_k + \frac12(\textbf{K}_1 + \textbf{K}_2),
\end{array}
$$
\end{small}
where $\Delta W_k \sim \mathcal{N}(0, \Delta t)$, and $S_k = \pm 1$, having each probability 1/2.

As described in the main text, the equation of motion of the center of mass of the levitated nanoparticle is \begin{small}
\begin{align*}
\dif x_t & = v_t \dif t ~,\\
m\dif  v_t& = -\nabla \Psi(x_t) \dif t -m\Gamma v_t \dif t + \sigma \dif W_t + mu(t)\dif t ~,
\end{align*}\end{small}
where 
$$
\Psi(x) = m\omega_0^2\left(\frac{x^2}{2}\right)
$$
and higher terms of the series expansion of the optical potential have not been considered.

The rest of the values needed to perform the simulations (i.e., $m$, $\Gamma$, $\sigma$, $\sigma_\xi$, and the electronic noise) have been calculated assuming:
\begin{itemize}
\item A temperature $T = 295$ K.
\item A spherical silica particle of radius $117.5$ nm and density 2200 kg/m$^3$.
\item $\gamma = m\Gamma$ follows Stoke's drag force, and is linear with the pressure for moderate levels of vacuum.
\item The noise intensity $\sigma$ satisfies the fluctuation-dissipation relation, i.e. $\sigma = \sqrt{2 k_B T m\Gamma}$.
\item $\sigma_\xi^2 \simeq 6 \times 10^{-24}$ m$^2$/Hz, the noise floor of our balanced detectors.
\item The Red Pitaya electronic noise approximately follows a normal distribution with $\sigma_\text{RP} = 1$ mV. The digital discretization noise has also been taken into account.
\end{itemize}

\subsection*{Mass determination}

We follow the procedure described in F. Ricci et. al.\citep{ricci2018accurate}, based on setting the number of elementary charges of the particle to a known value (we apply a high DC voltage $V_{\rm HV} \sim \pm 1~{\rm kV}$ to a bare electrode and create a corona discharge), driving the particle at a specific frequency $\omega_\text{dr}$ with a calibrated electric field and comparing the measurements of the CoM motion PSD to theory. 

Since the motion of the particle in the optical trap without any driving is purely thermal, its PSD $S_v(\omega)$ is well approximated by a Lorentzian function
$$
S_v(\omega) = \frac{\sigma^2/m^2}{(\omega^2 - \omega_0^2)^2 + \Gamma^2\omega^2}.
$$

From an experimental measurement of $S_v^{\text{th}}(\omega)$ we can extract the value of $S_v^{\text{th}}(\odr)$ and perform maximum likelihood estimation to obtain the values of $\omega_0$ and $\Gamma$ as fitting parameters. Introducing an electric driving, we determine the magnitude of the driven resonance $S_v(\odr)$ and calculate the electrical contribution  $S_v^{\rm el}(\odr) = S_v(\odr) - S_v^{\rm th}(\odr)$. 

The mass of the particle can ultimately be calculated considering the ratio $R_S=\frac{S_v^{\text{el}} (\odr)}{S_v^{\text{th}} (\odr)} =  \left. \frac{S_v-S_v^{\text{th}}}{S_v^{\text{ th}}}\right|_{\omega=\odr}$. Note that $S_v^{\text{el}} $ scales as $m^{-1}$ while  $S_v^{\text{th}}$ scales as $m^{-2}$. Thus, from their ratio we obtain:
\begin{equation} \label{eq:mass}
m = \frac{n_q^2 ~q_e^2 ~ E_0^2 ~ \mathcal{T} ~ }{8~ k_{\rm B} T ~ \Gamma ~ R_S}~,
\end{equation}
where $n_q$ is the number of elementary charges, $q_e$ the electron charge, $E_0$ the electric field amplitude, $\mathcal{T}$ the trace integration time, $k_B$ Boltzmann's constant, $\Gamma$ the damping and $R_S$ the previously calculated ratio.

\end{document}